\documentstyle[12pt]{ioplppt}
\input amssym.def
\input amssym

\def\be{\begin{equation}}
\def\ee{\end{equation}}
\def\beq{\begin{eqnarray}}
\def\eeq{\end{eqnarray}}

\begin{document}

\title{Proof of the cosmic no-hair conjecture for quadratic
homogeneous cosmologies}
\author{S Cotsakis\ftnote{1}{e-mail: skot@kerkis.math.aegean.gr},
J Miritzis\ftnote{2}{e-mail: imyr@kerkis.math.aegean.gr}}
\address{Department of Mathematics,
University of the Aegean,
Karlovassi
83200, Samos,
Greece}

\begin{abstract}
We prove the cosmic no-hair conjecture for all orthogonal Bianchi
cosmologies with matter in the $R+\beta R^2$ theory using the conformally
equivalent Einstein field equations, with the scalar field having the full
self--interacting potential, in the presence of the conformally related
matter fields. We show, in particular, that the Bianchi IX universe
asymptotically approaches de Sitter space provided that initially the scalar
three--curvature does not exceed the potential of the scalar field
associated with the conformal transformation. Our proof relies on rigorous
estimates of the possible bounds of the so-called Moss-Sahni function which
obeys certain differential inequalities and a non-trivial argument which
connects the behaviour of that function to evolution of the spatial part of
the scalar curvature.
\end{abstract}

\maketitle

\section{Introduction}

Attempts to tackle to isotropization problem in relativistic cosmology
date at least since the
pioneering work of Collins and Hawking \cite{CH} who proved isotropization
theorems for certain classes of orthogonal Bianchi spacetimes. A further
advance is the recent work by Heusler \cite{heus} which extended the
Collins--Hawking results to the case of a minimally coupled scalar field
matter source with a general convex potential. Interest in a particular
approach to the isotropization problem in cosmology renewed after the advent
of inflation as a mechanism for solving the problem and focused on proving
the so--called cosmic no--hair conjecture. This conjecture roughly speaking
states that general cosmological initial data sets when evolved through the
gravitational field equations are attracted (in a sense that can be made
precise) by the de Sitter space of inflation. In other words, if this
conjecture is true, inflation is a `transient' attractor of such sets, these
in turn quickly isotropize after the inflationary period and the
inflationary regime could thus be regarded as `natural' in view of its {\em %
prediction} of the (observed) large--scale homogeneity and isotropy of the
universe.

Higher derivative quantum corrections to the gravitational action of
classical general relativity are generally expected to play a significant
role at very high energies where a quantum gravitational field will
presumably dominate. It is not unreasonable to consider classical cosmology
in theories coming out of such nonlinear (higher derivative) gravitational
Lagrangians and in fact, one expects that there exist close links between
properties of such ``higher derivative cosmologies'' and those of general
relativistic cosmology. It is obvious that the resolution of aspects of the
singularity, isotropization and recollapse problems of cosmological spacetimes
is of paramount
importance also in this extended framework.

In this paper, we prove the cosmic no--hair conjecture for all Bianchi
(including type IX) matter cosmologies which are derived from a quadratic
action in the scalar curvature that is of the form $L=R+\beta R^2+L_m$ where 
$L_m$ denotes the matter Lagrangian.

Other aspects of this problem were considered previously, for instance, for the
case of the Einstein field equations with a cosmological constant by Wald 
\cite{wald2}, and Maeda \cite{maed} for the case of the quadratic theory $%
L=R+\beta R^2$ in vacuum for all Bianchi models except the IX. More
specifically, Wald \cite{wald2} proved, in the particular case of a true
cosmological constant, that this conjecture is true for all Bianchi models,
except possibly for type IX . Wald's proof served as the prototype for most
subsequent works where, either a true cosmological constant was assumed \cite
{jess} or the validity of the cosmic no--hair conjecture was tested in the
context of particular models: chaotic inflation \cite{mosa,heus}, power-law
inflation \cite{ihc,kima}, inflation due to a $R^2$ theory \cite
{maed,cofl,berk} or in more general $f\left( R\right) $ theories \cite{mico}.

Our results contribute to an extension of previous works in two
ways. We provide a detailed study of the role that matter plays in the
problem and treat rigorously the important case of the Bianchi IX model
(including matter fields) {\em in the case of the full self-interacting
potential} arising from the conformal transformation. Our proof relies on
the use of an energy-like function which we call the Moss-Sahni function and
is accomplished through a detailed study of energy estimates of the
associated quantities that obey certain differential inequalities. We show
that once the required bound for the Moss-Sahni function is obtained, the
spatial part of the scalar curvature asymptotically approaches zero almost
exponentially.

The field equations take the form 
\begin{equation}
R_{ab}-\frac 12g_{ab}R-\frac \beta {1+2\beta R}\left( 2\nabla _a\nabla
_bR-2g_{ab}\Box R+\frac 12g_{ab}R^2\right) =T_{ab}\left( {\bf g}\right) .
\label{quae}
\end{equation}
Under a conformal transformation of the metric 
\begin{equation}
\widetilde{g}_{ab}=\left( 1+2\beta R\right) g_{ab},  \label{conf}
\end{equation}
with 
\begin{equation}
\varphi =\sqrt{\frac 32}\ln \left( 1+2\beta R\right) ,  \label{qusf}
\end{equation}
the field equations (\ref{quae}) become the Einstein equations in the new
space-time $\left( M,\widetilde{{\bf g}}\right) $ 
\begin{equation}
\widetilde{R}_{ab}-\frac 12\widetilde{g}_{ab}\,\widetilde{\,R}=\nabla
_a\varphi \nabla _b\varphi -\frac 12\widetilde{g}_{ab}{}\,\left( \nabla
_c\varphi \nabla ^c\varphi \right) -\widetilde{g}_{ab}V+\widetilde{T}%
_{ab}\left( \widetilde{{\bf g}}\right) ,  \label{coqe}
\end{equation}
\begin{equation}
\widetilde{\stackrel{}{\Box }}\varphi -V^{^{\prime }}\left( \varphi \right)
=0,  \label{box}
\end{equation}
where the potential $V$ is \cite{baco}, \cite{maed} 
\begin{equation}
V=\frac 1{8\beta }\left[ 1-\exp \left( -\sqrt{\frac 23}\varphi \right)
\right] ^2.  \label{pote}
\end{equation}
As Maeda \cite{maed} has pointed out, this potential has a long and flat
plateau. When $\varphi $ is far from the minimum of the potential, $V$ is
almost constant $V_\infty \equiv \lim_{\varphi \rightarrow +\infty }V\left(
\varphi \right) =1/\left( 8\beta \right) .$ Thus $V$ has the general
properties for inflation to commence and $V_\infty $ behaves as a
cosmological term. The constant $\beta $ is of order $10^{14}\;l_{PL}^2$ 
\cite{mms}.

The plan of the paper is as follows. In the next Section we prove the cosmic
no-hair conjecture for all Bianchi universes except IX, while Section 3 is
devoted to the analysis of the Bianchi IX model. Precise statements of the
results proved may be found in those Sections. We conclude by pointing out
how our results could lead to the treatment of several more general cases.

\section{All Bianchi models except type IX}

In this Section we show using Einstein's equations (\ref{coqe}) and the
equation of motion of the scalar field (\ref{box}), that, {\em all Bianchi
models which are initially on the flat plateau of the potential (\ref{pote}%
), except probably Bianchi IX, with a matter content satisfying the strong
and dominant energy conditions, rapidly approach de Sitter space-time. }This
is established for the full scalar field potential arising from the
conformal transformation.

We shall work exclusively in the conformal picture and so for simplicity we
drop the tilde. Choose $n^a$ to be the unit geodesic vector field, normal to
the homogeneous hypersurfaces. The spatial metric is related to the
space-time metric as $h_{ab}=g_{ab}+n_an_b$ and $n^a\nabla _a=\partial
/\partial t,$ where $t$ denotes proper time along the integral curves of $%
n^a.$ The scalar curvature $R$ becomes a function of time, hence the scalar
field introduced by eq. (\ref{qusf}) is homogeneous. Ordinary matter is
assumed to satisfy the strong and dominant energy conditions, namely, 
\begin{equation}
\begin{array}{lll}
T_s & \equiv & \left( T_{ab}-\frac{1}{2}Tg_{ab}\right) n^an^b\geq 0 \\ 
T_d & \equiv & T_{ab}u^a u^b \geq 0,\hspace{0.5cm}T^a _b u^b \hspace{0.3cm}%
{\rm {non-spacelike,}}
\end{array}
\label{deco}
\end{equation}
for any unit timelike vector field $u^a.$

We make use of the time-time component of Einstein's equations (\ref{coqe}) 
\begin{equation}
\frac 13K^2=\sigma ^2+T_d+\frac 12\dot{\varphi}^2+V-\frac 12\;^{\left(
3\right) }R  \label{ooco}
\end{equation}
and the Raychaudhuri equation 
\begin{equation}
\stackrel{.}{K}=-\frac 13K^2-2\sigma ^2-T_s-\stackrel{.}{\varphi }^2+V.
\label{rayc}
\end{equation}
The equation of motion for the scalar field (\ref{box}) becomes 
\begin{equation}
\ddot{\varphi}+K\dot{\varphi}+V^{\prime }\left( \varphi \right) =0.
\label{emsf}
\end{equation}
Here $K$ is the trace of the extrinsic curvature $K_{ab}$ of the homogeneous
hypersurfaces and is related to the determinant $h$ of the spatial metric by 
\begin{equation}
K=\frac d{dt}\left( \ln h^{1/2}\right) .  \label{expa}
\end{equation}
We note that in \cite{kima,ha}, the system above is reduced to a two
dimensional autonomous form as a consequence of the exponential form of the
potential which we do not assume. Our case is more involved since the system
cannot be reduced further and so we have to proceed in a different way.
We make the following assumptions:
\begin{itemize}
\item[1.] The initial value $\varphi _i$ of the scalar field is large and
positive (i.e., the universe is on the flat plateau of the potential).
\item[2.] The kinetic energy of the field is negligible compared to the potential
energy.
\item [3.] The universe is initially expanding i.e., $K>0$ at some arbitrary
time $t_i$.
\end{itemize}
With  assumption (3), Eq. (\ref{ooco}) implies that the universe will expand for all
subsequent times i.e., $K>0$ for $t\geq t_i$ for all Bianchi models, except
possibly type-IX. Defining the energy density of the scalar field by $%
E\equiv \frac 12\stackrel{.}{\varphi }^2+V$, we find by (\ref{emsf}) that 
\begin{equation}
\dot{E}=-K\dot{\varphi}^2.
\end{equation}
Hence, in an expanding universe the field looses energy and slowly rolls
down the potential. The ``effective'' regime of inflation is the phase
during the time interval $t_f-t_i$ needed for the scalar field to evolve
from its initial value $\varphi _i$ to a smaller value $\varphi _f,$ where $%
\varphi _f$ is determined by the condition that $V\left( \varphi _f\right)
\simeq \eta V_\infty .$ The numerical factor $\eta $ is of order say $0.9,$
but its precise value is irrelevant. As we shall show, the universe becomes
de Sitter space during the effective regime followed by the usual FRW model
when the cosmological term vanishes. To this end, we define a function $S$,
which plays the same role as in Moss and Sahni \cite{mosa}, by 
\begin{equation}
S=\frac 13K^2-E.  \label{s}
\end{equation}
In all Bianchi models except IX, using (8) we see that this function is
non-negative due to the dominant energy condition and the fact that in these
models the scalar spatial curvature is non-positive. Taking the time
derivative of $S$ and using eqs. (\ref{ooco}) and (\ref{rayc}) we obtain 
\begin{equation}
\dot{S}=-\frac 23KS-\frac 23\left( 2\sigma ^2+T_s\right) K.
\end{equation}
It follows that $\dot{S}\leq -(2/3)KS$ or, 
\begin{equation}
\dot{S}\leq -\frac 23S\sqrt{3\left( S+E\right) }.  \label{sder}
\end{equation}
This differential inequality cannot be integrated immediately because $E$ is
a function of time (albeit slowly-varying). However as is well known \cite
{lind}, in order to have inflation, $E$ must be bounded below and in this case
assumption (1) above implies that we may  assume
 that the scalar field is large enough and so, without loss of
generality, we may set $E\geq \eta V_\infty $. The choice of $\eta $
affects only slightly the isotropization time, but the qualitative behavior
of the model is the same. Therefore, inequality (\ref{sder}) implies that 
\begin{equation}
S\leq \frac{3m^2}{\sinh ^2\left( mt\right) },\hspace{0.5cm}m=\sqrt{\eta
V_\infty /3}.  \label{upbo}
\end{equation}
From eq. (\ref{ooco}) we see that, as $t$ increases, the shear,
three-curvature and energy density of matter rapidly approach zero and
because of the dominant energy condition, all components of the
energy-momentum tensor approach zero. It follows that the universe
isotropizes within one Hubble time ($1/\sqrt{V_\infty }\sim 10^7\;t_{PL}$).

\section{Bianchi type IX models}

In this Section we show that {\em Bianchi-type IX also isotropizes if
initially the scalar three-curvature }$^{\left( 3\right) }R${\em \ is less
than the potential }$V${\em \ of the scalar field.} In the present case, the
scalar curvature $^{\left( 3\right) }R$ may be positive and the argument of
the previous Section does not apply directly. In particular, the function $S$
is not bounded below from zero.\footnote{%
However, as Wald \cite{wald2} points out, for some non-highly positively
curved models premature recollapse may be avoided provided a large positive
cosmological constant compensates the $-\frac 12\,^{\left( 3\right) }R$ term
in eq. (\ref{ooco}). In our case, of course, it is the potential $V\left(
\varphi \right) $ which acts as a `cosmological term'. A similar proof to
that of Wald \cite{wald2} for the Bianchi IX model, was given by Kitada and
Maeda \cite{kima} in the case of an exponential potential leading to
power-law inflation.} However, we can estimate an upper bound for $S$ by
observing that, either $S$ is bounded above from zero, or, if $S$ is initially
positive, inequality (\ref{sder}) implies as before that an upper bound for $%
S$ is given by (\ref{upbo}). Collectively, we have the overall result that 
\begin{equation}
S\leq \max \left\{ 0,\;3m^2\sinh ^{-2}\left( mt\right) \right\} .
\label{upper}
\end{equation}
(Had we chosen an exponential potential, this bound would have had the form
found in \cite{kima}, p.1418). An estimation of a lower bound for $S$
derives from the fact that the largest positive value the spatial curvature
can achieve is determined by the determinant of the three metric 
\begin{equation}
^{\left( 3\right) }R_{\max }\propto h^{-1/3}\equiv \exp \left( -2\alpha
\right) .  \label{rmax}
\end{equation}
We obtain a lower bound for $S$ in the following way. We assume that the
three-scalar curvature $^{\left( 3\right) }R$ is less than the potential
energy since otherwise it is known that premature recollapse commences
before vacuum domination drives the universe to inflate \cite{barr}. Then,
eq. (\ref{ooco}) immediately yields 
\[
\frac 13K^2-\frac 12\stackrel{.}{\varphi }^2\geq 0. 
\]
Observe that for any $0<\lambda <\sqrt{2/3}$ the above inequality implies
that for $K>0,$ 
\begin{equation}
\frac 13K-\frac 12\lambda \stackrel{.}{\varphi }\geq \frac 13K-\frac 1{\sqrt{%
6}}\,\left| \stackrel{.}{\varphi }\right| \geq 0.  \label{tric}
\end{equation}
Suppose that initially i.e.{\it ,} at time $t_i$ we have $V>\,^{\left(
3\right) }R_{\max }$. We claim that this inequality holds during the whole
period of the effective regime of inflation. Following \cite{kima} we define
a function $f$ by 
\begin{equation}
f\left( t\right) \equiv \ln \frac V{^{\left( 3\right) }R_{\max }},\;\;t\in
\left[ t_i,t_f\right] ,  \label{f}
\end{equation}
whose initial value is positive by the above assumption. Differentiating we
find 
\begin{equation}
\stackrel{.}{f}=\frac{\stackrel{.}{V}}V+2\stackrel{.}{\alpha }=2\left[ \frac{%
\exp \left( -\sqrt{\frac 23}\varphi \right) }{1-\exp \left( -\sqrt{\frac 23}%
\varphi \right) }\sqrt{\frac 23}\stackrel{.}{\varphi }+\frac 13K\right]
\label{fder}
\end{equation}
since $K=3\stackrel{.}{\alpha },$ by eq. (\ref{expa}). During the effective
regime of inflation $V\left( \varphi \right) \geq \eta V_\infty ,$ hence
solving for $\varphi $ this inequality we immediately verify that the
coefficient of $\dot{\varphi}$ in the brackets is less than $1/\sqrt{6}.$
Therefore, inequality (\ref{tric}) implies that $\dot{f}\geq 0$ . We
conclude that, if $K>0$ and $f\left( t_i\right) >0$ initially, then for $%
t\in \left[ t_i,t_f\right] $ we obtain $f\left( t\right) \geq f\left(
t_i\right) $ and our assertion follows.

We are now in a position to estimate the required bound for $S$. First, from
eq. (\ref{ooco}) it is evident that 
\begin{equation}
-\frac 12\;^{\left( 3\right) }R_{\max }\leq S  \label{20}
\end{equation}
and since $V/2>\,^{\left( 3\right) }R_{\max }/2,$ eq. (\ref{ooco}) again
implies that 
\begin{equation}
K\geq \sqrt{\frac 32V}>\sqrt{\frac 32\eta V_\infty }.
\end{equation}
Remembering that $K=3\dot{\alpha }$ and using the last inequality and eq. (%
\ref{rmax}), we see that $^{\left( 3\right) }R_{\max }$ decays faster than $%
\exp \left[ -\sqrt{2\eta V_\infty /3}\left( t-t_i\right) \right] $.
Therefore (\ref{20}) becomes 
\begin{equation}
-\frac 12\,^{\left( 3\right) }R_{\max }\left( t_i\right) \exp \left[ -\sqrt{%
\frac 23\eta V_\infty }\left( t-t_i\right) \right] \leq S.  \label{lobo}
\end{equation}
Combining this lower bound for $S$ with the upper bound (\ref{upper}), we
conclude that $S$ vanishes almost exponentially. From eq. (\ref{ooco}), $%
-2S\leq \,^{\left( 3\right) }R\leq \,^{\left( 3\right) }R_{\max }$ and
therefore $^{\left( 3\right) }R$ damps to zero just as $S$ and $^{\left(
3\right) }R_{\max }.$

We conclude that when the universe is initially on the plateau, the shear,
the scalar three-curvature and all components of the stress-energy tensor
approach zero almost exponentially fast with a time constant of order $\sim
1/\sqrt{V_\infty }.$\footnote{%
Actually, the time constant for isotropization in type IX is longer by $%
\sqrt{2}$ than in other types. The situation is similar to that encountered
in Kitada and Maeda \cite{kima}.}

We discussed inflation in the equivalent space-time $\left( M,\widetilde{%
{\bf g}}\right) ,$ but it is not obvious that the above attractor property
is maintained in the original space-time $\left( M,{\bf g}\right) .$ This is
probably an unimportant question since there is much evidence that in most
relevant cases the rescaled metric $\widetilde{{\bf g}}$ is the real
physical metric \cite{cots2}. However, Maeda has pointed out \cite{maed}
that inflation also naturally occurs in the original picture since during
inflation the scalar field changes very slowly and the two metrics are
related by $\widetilde{{\bf g}}= \exp\left( \sqrt{\frac{2}{3}}\varphi\right) 
{\bf g}$.

We now move on to show that{\em \ the time needed for the potential energy
to reach its minimum is much larger than the time of isotropization.}
Therefore, the universe reaches the potential minimum $\left( \varphi
=0\right) $ at which the cosmological term vanishes and consequently evolves
according to the standard Friedmann cosmology. Wald's proof assumes the
existence of a cosmological constant. However, in realistic inflationary
models the universe does not have a true cosmological constant but rather a
vacuum energy density which during the slow evolution of the scalar field
behaves like a cosmological term which eventually vaniches. Therefore we are
faced with the question of whether or not the universe evolves towards a de
Sitter type state {\em before} the potential energy of the scalar field
reaches its minimum. In any consistent no-hair theorem one has to verify
that the time necessary for isotropization is small compared to the time the
field reaches the minimum of the potential. The following argument shows
that in our case the vacuum energy is not exhausted before the universe is
completely isotropized. (Similar estimates were derived in \cite{mosa} for
the case of the standard quadratic potential.) Imagine that at the beginning
of inflation the universe is on the flat plateau of the potential. It is
evident that the time $t_f-t_i$ needed for the scalar field to evolve from
its initial value $\varphi _i$ to a smaller value $\varphi _f$ is smaller in
the absence of damping than that in its presence. In the absence of damping, 
$t_f-t_i$ is easily obtained from integrating the equation of motion of the
scalar field (\ref{emsf}) ($K=0).$ Taking for example a $\varphi _f$ such
that $V\left( \varphi _f\right) \simeq \eta V_\infty $ one finds that $%
t_f-t_i$ is more than $65$ times the time $\tau $ of isotropization. The
presence of damping increases the time interval $t_f-t_i$ and a larger
anisotropy damps more efficiently the slow rolling of the scalar field, thus
producing more inflation. It follows that when due account of damping is
taken, the period of the effective regime of inflation is more than
sufficient for the complete isotropization of the universe.

\section{Conclusions}

It is known \cite{cofl} that the quasi-exponential solution of the $R+\beta
R^2$ theory considered in this paper is an attractor to all isotropic
solutions of this theory. Our analysis here suggests that the solution could
be shown to attract the class of orthogonal Bianchi universes as well. If
this is indeed the case, it would be interesting to ask whether this
attractor is unique in the space of all higher order gravity theories. One
way to directly verify this might be through a perturbation analysis and the
corresponding investigation of the asymptotic structure of the solutions.

We believe that the proof of the cosmic no--hair conjecture presented here
could be extended with adjustments in two directions namely, in an arbitrary
number of dimensions and also  to the class of titled Bianchi
models. In fact, such a demonstration in some titled models
would amount to a first test of this
conjecture in cases of some inhomogeneity. An analysis along these lines
might be more tractable than say attacking directly a genuine inhomogeneous
case such as for instance that of $G_2$ cosmologies wherein the dynamics is
described by systems of partial differential equations. Also a more direct
analysis along the lines adopted here could lead to an extension of our
results in higher dimensions.

{\bf Acknowledgements} We are grateful to J.D. Barrow and
G. Flessas for many useful and critical comments which
greatly improved the present work. One of us (S.C.) received support from
the Research Commission of the University of the Aegean (Grant Numb.
3516075-2) and the General Secretariat for Science and Technology (Grant
Numb. 1361/4.1) which is gratefully acknowledged.

\end{document}